\documentclass[useAMS,usenatbib]{pasa}
\usepackage{epsfig}

\title[A Second Shell in the Fornax dSph Galaxy]{A Second Shell in the Fornax dSph Galaxy}
\author[M.~G.~Coleman \& G.~S.~Da Costa]{M.~G.~Coleman \& G.~S.~Da Costa}
\affil{
Research School of Astronomy \& Astrophysics\\
Institute of Advanced Studies, The Australian National University\\
Cotter Road, Weston Creek, ACT 2611, Australia\\
coleman@mso.anu.edu.au}

\begin{document}

\maketitle

\label{firstpage}

\begin{abstract}
In the search for tidal structure in Galactic satellite systems, we have conducted a photometric survey over a 10 deg${}^2$ area centred on the Fornax dSph galaxy.  The survey was made in two colours, and the resulting colour-magnitude data were used as a mask to select candidate Fornax RGB stars, thereby increasing the contrast of Fornax stars to background sources in the outer regions.  Previously, we reported the presence of a shell (age 2 Gyr) located towards the centre of Fornax.  In this contribution we reveal a second shell, significantly larger than the first, located $1.3^{\circ}$ NW from the centre of Fornax, outside the nominal tidal radius.  Moreover, the distribution of Fornax RGB stars reveals two lobes extending to the spatial limit of our survey, and aligned with the minor axis and with the two shells.  These results support the hypothesis of a merger between Fornax and a gas-rich companion approximately 2 Gyr ago.
\end{abstract}


\section{Introduction}
Observations of large-scale structure throughout the Universe indicate galaxies are arranged in clusters and filaments, surrounding large voids which contain few luminous objects.  The processes leading to this structure have been investigated by dark matter-dominated cosmological simulations (for example, \citealt{nfw96,nfw97,moore99}) which predict that a complex system of hierarchical merging formed the galaxies of the present day.  Evidence of these mergers can be seen as remnant structure in large spiral and elliptical galaxies.  \citet{malin80} detected phase-wrapped shells in the haloes of large galaxies through a process of photographic amplification \citep{malin78}, and found that such structure is relatively common \citep{malin83}.  Simulations by \citet{hernquist88,hernquist89} indicate that these shells originated through the merging of a smaller companion galaxy with its host object, with some of the companion's mass ($\sim$$10\%$) distributed as shells moving on radial orbits through the potential of the large galaxy.  Hence shell structure is a common remnant of large galaxy mergers.

If such structure can be seen in large galaxies, it is reasonable to expect a similar effect in smaller bodies such as the dwarf galaxies.  CDM simulations predict that these objects have also formed through a process of hierarchical merging, and are therefore a compilation of several clumps of matter.  Also, the process of galaxy-galaxy interactions are thought to be scale-free; hence the formation of shell structure at large galactic scales should be echoed at the scale of dwarf galaxies.  However, by redshift $z \sim 10$ dwarf galaxies are expected to have formed through the merger of these small dark halos, and late infall in a typical dwarf halo is thought to be negligible (Power 2003, personal communication).  Thus, a typical dwarf galaxy is not expected to display recently formed shell structure.

Based on a deep photometric dataset covering the inner 1/3 deg${}^2$ of the Fornax dSph obtained by \citet{stetson98}, we previously reported the discovery of a shell-like feature located 17 arcmin ($\sim$1.8 core radii) from the centre of the system (\citealt{coleman04a}; hereafter C04).  The analysis revealed that the clump is composed almost exclusively of a stellar population with an age of 2 Gyr.  It is aligned with the major axis of Fornax, and is situated on the minor axis.  Our inference was that the stellar association represents shell structure in a dwarf galaxy, the remnant of a merger between Fornax and a gas-rich companion approximately 2 Gyr ago.  However, such a conclusion was based on a single shell located towards the central regions of Fornax.

In this contribution we report the discovery of a second shell, located outside the nominal tidal radius of Fornax, based on a wide-field photometric survey of the dSph.  This shell is oriented parallel to the original structure described in C04, and is located on the opposite side of Fornax.  In addition, two large structures aligned with these shells extend approximately two tidal radii from the centre of Fornax.  Based on these results, the system appears to have experienced a significant interaction with a gas-rich companion in the recent past.  A full description of the Fornax survey and analysis techniques can be found in \citet{coleman04b}.

\section{The Fornax Survey}
\begin{figure}
\begin{center}
\includegraphics[width=8cm]{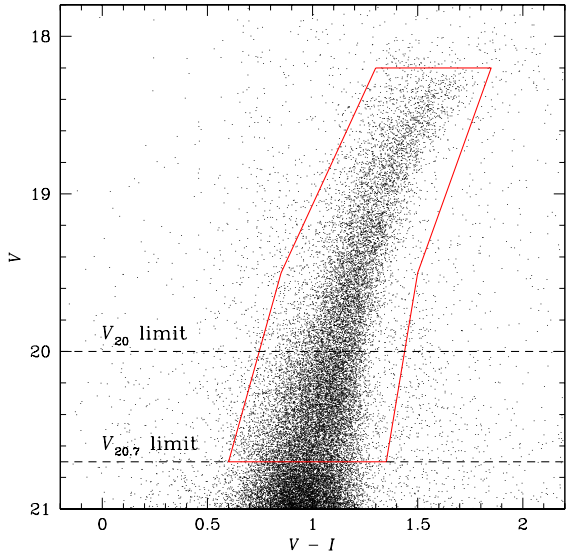}
\caption{The colour-magnitude data from the inner $40'$ of Fornax, compiled from our photometry survey.  The red lines outline the selection region for candidate Fornax stars.  The nominal photometric limit at $V=20$ represents the limit of the entire dataset, while the deep photometric limit ($V=20.7$) applies for six of the sixteen fields. \label{fornaxcmd}}
\end{center}
\end{figure}

We have collected photometry in $V$ and $I$ bands for an area of $3.1 \times 3.1$ deg${}^2$ centred on the Fornax dSph using the Wide-Field Imager (WFI) attached to the SSO 1 metre telescope.  The region was imaged using a mosaic of $4 \times 4$ fields (labelled F1, \dots, F16), and the data are complete to a magnitude of $V=20$ at the colour of the RGB.  All fields contain an overlap region (width $\sim$$5'$) with each adjacent field, which we used to measure the internal accuracy of the photometry and astrometry.  These analyses, as well as the coordinates and a map of the survey region are presented in \citet{coleman04b}.  The colour-magnitude diagram (CMD) from the inner $40'$ of Fornax is shown in  Fig.\ \ref{fornaxcmd}, where the upper dotted line illustrates the completeness limit of the survey.  By selecting stars within the RGB selection range, we are able to increase the `signal' of Fornax stars compared to the background field population (the `noise').  We selected candidate RGB stars down to the completeness limit, and label this the `$V_{20}$' dataset.  However, six of the sixteen fields we imaged were complete to a deeper limit of $V=20.7$, represented as the lower dotted line in Fig.\ \ref{fornaxcmd}.  Hence we selected stars down to this deeper limit in these six fields to further increase the signal-to-noise of Fornax stars, and label this the `$V_{20.7}$' dataset.  A full discussion of both datasets is presented in \citet{coleman04b}.

\begin{figure}
\begin{center}
\includegraphics[width=8cm]{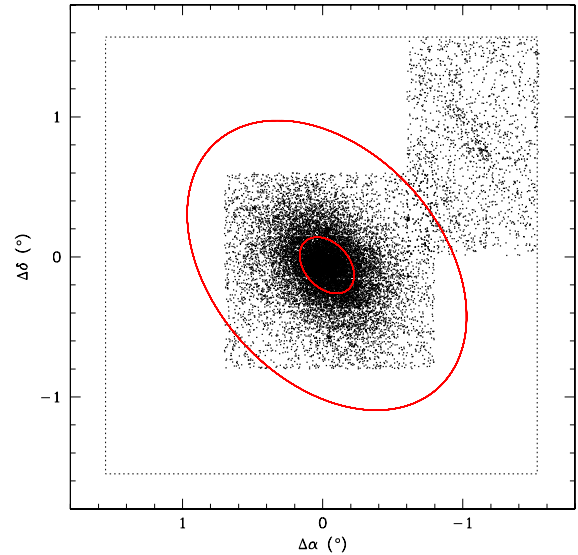}
\caption{Distribution of stars in the $V_{20.7}$ dataset, which only includes six of the sixteen fields imaged.  The full survey area is marked as the dotted line, where the remaining ten fields have been excised from this image.  The points plotted are candidate RGB stars for the six fields with complete photometry down the the deep photometric limit displayed in Fig.\ \ref{fornaxcmd}.  The red ellipses define the core and tidal radii from \citet{m98}.  The original shell described in C04 is located at $\Delta \alpha = 0.10^{\circ}$, $\Delta \delta = -0.25^{\circ}$ \label{fornaxrgbxy}}
\end{center}
\end{figure}

In Fig.\ \ref{fornaxrgbxy} we show the distribution of the stars in the $V_{20.7}$ dataset.  The outer red ellipse defines the nominal tidal radius of this system, taken from \citet{m98}.  Perhaps the most apparent feature in Fig.\ \ref{fornaxrgbxy} is a region of increased stellar density located approximately $1.3^{\circ}$ north-west from the centre of Fornax.  This apparent overdensity is situated in the field F4 {\em outside} the nominal tidal radius.  An inspection of the CCD images for F4 indicates all the stars within this feature are real stellar sources, and the star finding algorithm does not appear to have missed any stars in the remainder of the field.  Also, given the high Galactic latitude of Fornax, we do not expect this feature to be caused by dust extinction.  Correspondingly, a visual inspection of the IRAS $100 \mu$m map \citep{schlegel98} does not show any significant dust structure in the region of this feature.

\begin{figure}
\begin{center}
\includegraphics[width=8cm]{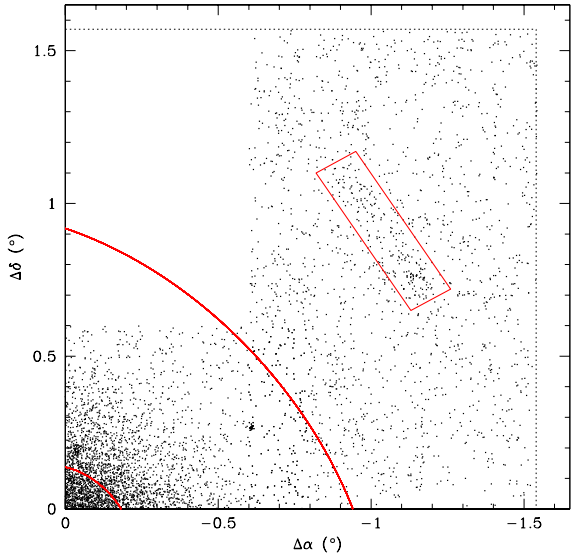}
\caption{Distribution of stars in the $V_{20.7}$ dataset for the NW quadrant only.  The red box outlines the shell selection region. \label{fornaxrgbnwxy}}
\end{center}
\end{figure}

Thus, the feature merits further analysis.  We have measured the total surface density in the vicinity of the feature to be $1.7 \pm 0.1$ times that of the remainder of F4, where the region we have selected is outlined in Fig.\ \ref{fornaxrgbnwxy}.  That is, at least one third of the stars in the overdense region are part of the feature, while the remainder form the background population.  We removed the background population \citet{coleman04b} and, assuming the feature lies at the same distance modulus as Fornax [$(m-M)=20.70$; \citealt{m98}], then it has an integrated absolute magnitude of $M_V=-6.7 \pm 0.2$, comparable to the five globular clusters associated with Fornax.  A visual inspection of the colour-magnitude data for all stars in this feature \citep{coleman04b} reveals that the stellar population defines an RGB qualitatively matching that of Fornax.

\section{Analysis}
\begin{figure}
\begin{center}
\includegraphics[width=8cm]{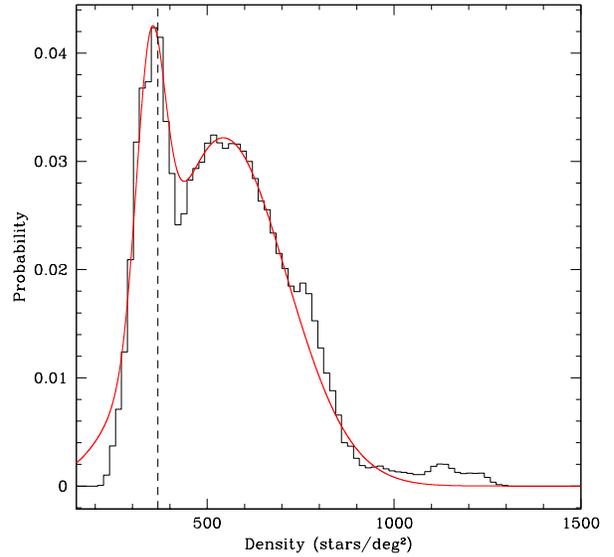}
\caption{Monte Carlo simulation of the density distribution of Fornax RGB-selected stars, showing the probability of measuring a given density beyond the nominal tidal radius.  The dashed line corresponds to the (normalised) \citet{rat85} star counts.  The red line indicates the sum of two best-fit Gaussians to the low and medium density background populations. \label{mccontour}}
\end{center}
\end{figure}

A full description of the analysis techniques are presented in \citet{coleman04b}.  Here we describe one of the tools used to statistically investigate the extra-tidal stellar structure.  We constructed a Monte Carlo simulation where each iteration consisted of placing a circle (radius 12 arcmin) at a random position {\em outside} the nominal tidal radius and counting the number of stars therein.  By completing $10^6$ iterations, we were able to construct a function describing the probability of measuring a given stellar density beyond the Fornax tidal limit.  In order to completely sample the survey area, this simulation was conducted over the $V_{20}$ dataset.  The star counts within circles falling partially outside the survey region were scaled by the appropriate amount to correct for the non-surveyed area.  The resulting probability function is displayed in Fig.\ \ref{mccontour}.  A random distribution of stars would be expected to display a normal probability distribution in density.  However, immediately visible in Fig.\ \ref{mccontour} is a triple-peaked structure, thus implying the extra-tidal region consists of three populations with distinguishable densities.  The dotted line corresponds to the predicted density of the field star population, determined by \citet{rat85} using the \citet{bahcall80} Galactic model.  \citet{rat85} estimate the star count uncertainty at $25\%$, hence we conclude the low-density peak corresponds to the true field star density.  We measured the standard deviation from the mean density of the field population, and this value ($\sigma_{\mbox{\scriptsize f}}$) allows us to measure the statistical significance of the higher density structures.

%\begin{figure}
%\begin{center}
%\includegraphics[width=8cm]{fornaxcmd_rgb_deep_sel.eps}
%\caption{The red points are all stars situated in the overdense region, while the black points are the CMD data displayed in Fig.\ \ref{fornaxcmd}. \label{clumpcmd}}
%\end{center}
%\end{figure}

The high-density peak (centred at a density of 1150 stars arcmin${}^{-2}$) corresponds to the feature described in Sec.\ 2, situated at $\Delta \alpha = -1.0^{\circ}$, $\Delta \delta = 0.8^{\circ}$ in Fig.\ \ref{fornaxrgbxy}.  This appears to be the only high-density structure beyond the nominal tidal radius of Fornax.  With a mean stellar density $\sim$$20\sigma_{\mbox{\scriptsize f}}$ higher than that of the field population, the feature is clearly separated from the background.  It is located on the opposite side of Fornax to the previous shell (C04), and both are situated on the minor axis and aligned parallel to the major axis.  Therefore this structure appears to be a second Fornax shell.

\begin{figure}
\begin{center}
\includegraphics[width=8cm]{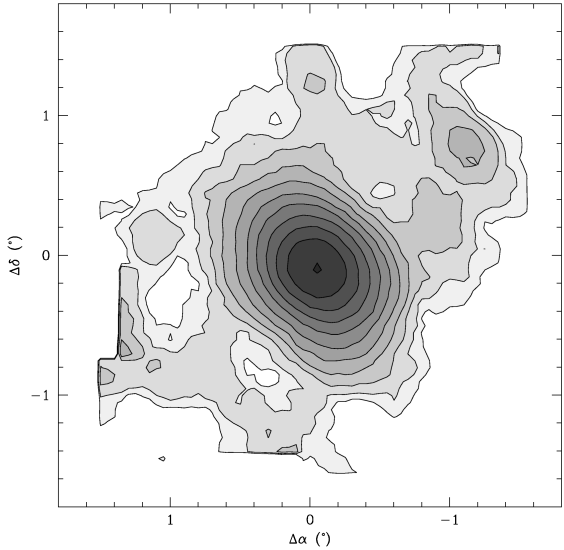}
\caption{Distribution of Fornax RGB stars in the $V_{20}$ dataset where each star has been convolved with a Gaussian of approximate width $10'$.  The contours are logarithmically spaced, and are fitted with a smoothing length of $3.0'$.  The first contour represents a stellar density $4.5\sigma_{\mbox{\scriptsize f}}$ above the field star population. \label{fornaxcontour}}
\end{center}
\end{figure}

Finally, we examine the medium-density population.  The contour plot in Fig.\ \ref{fornaxcontour} was constructed by convolving each star in the $V_{20}$ dataset with a Gaussian of approximate width $10'$.  The contour smoothing length is 3 arcmin.  The lower significance limit is $4.5\sigma_{\mbox{\scriptsize f}}$, hence the first contour traces all structure with a density $4.5\sigma_{\mbox{\scriptsize f}}$ above that of the field star population.  This contour outlines two `lobes' extending beyond the nominal tidal radius of Fornax, located on the minor axis.  Assuming they are at the same distance as Fornax, with a similar stellar population, they measure $\sim$8 kpc in length with an integrated absolute visual magnitude of $M_V \sim -9$.

In summary, beyond the nominal tidal radius of the dSph, the NE/SW quadrants of the survey contain only the field star population, with no discernible excess of Fornax RGB stars.  However, in the NW/SE quadrants, we detected two large structures extending at least 1.0 deg beyond the tidal limit, and these appear symmetrical along the minor axis of Fornax.  Moreover, we have located a second shell approximately $1.3^{\circ}$ NW of the centre of Fornax.

\section{Discussion}
In C04 we proposed the existence of shell structure in Fornax based on a single feature located approximately two core radii from the centre of the system.  Such structure would arise from the interaction between Fornax and a companion object.  Another possibility is that the two large lobes of material extending to either side of the dSph represent tidal tails.  They qualitatively match simulations of tidal tails \citep{helmi01,mayer01} which are caused by interaction with the Milky Way.  These models also indicate that material leaves the satellite in bursts, which may produce the observed shell-like structure.

However, clear tidal tails have only been observed in a single dSph satellite of the Galaxy; the Sagittarius dSph.  The Sagittarius orbit brings it within approximately 15 kpc from the Galactic centre with an orbital eccentricity of $e \approx 0.75$ \citep{law04}, and thus Sgr experiences substantial tidal forces.  In contrast, Fornax is located at a distance of 140 kpc \citep{m98} and proper motion measurements indicate it is on a polar orbit with a small eccentricity ($e=0.27$) and is currently at perigalacticon \citep{dinescu04}.  Moreover, Fornax contains a mass larger than all the other Galactic dSphs combined (excluding Sagittarius), which further precludes the possibility of stars escaping from this relatively deep potential well.  Thus, given the relatively large mass, distance and orbital parameters of Fornax, we do not expect to observe such clear evidence of Milky Way interaction.

Further, the inner shell described in C04 demonstrated a clear signature; it was dominated by stellar population with an age of 2 Gyr, significantly different to the mixed stellar population observed in Fornax \citep{stetson98}.  If the observed substructure is due to the gravitational influence of the Galaxy, then it would contain the same stellar population as the remainder of the system.

The high density structure located $1.3^{\circ}$ NW of Fornax is clearly shell-like in appearance.  In the case of shell structure, the lobes would represent those portions of shell for which we do not see a limb brightening effect; in essence, they lie in front of, or behind, the Fornax system.  These appear remarkably similar to the simulations of interacting dark halos presented by \citet{knebe04}.  To further investigate the shell possibility, we will obtain deep photometry of the shell region.  If the stellar population is similar to the Fornax system, this would indicate the substructures are tidal tails.  However, if the stars in the shell display a marked age difference from those in Fornax, this would support the possibility of shell structure in a dSph.

%Uncomment if needed
%\section*{Appendix}

\label{lastpage}


\begin{thebibliography}{}

\bibitem[Bahcall \& Soneira(1980)]{bahcall80} Bahcall, J.~N.~\& Soneira, R.~M.\ 1980, ApJS, 44, 73 

\bibitem[Coleman et al.(2004)]{coleman04a} Coleman, M., Da Costa, G.~S., Bland-Hawthorn, J., Mart\'{\i}nez-Delgado, D., Freeman, K.~C., \& Malin, D.\ 2004, AJ, 127, 832 (C04)

\bibitem[Coleman et al.(in prep.)]{coleman04b} Coleman, M., Da Costa, G.~S., Bland-Hawthorn, \& J., Freeman, K.~C., in prep.

\bibitem[Dinescu et al.(2004)]{dinescu04} Dinescu, D.~I., Keeney, B.~A., Majewski, S.~R., \& Girard, T.~M.\ 2004, AJ, 128, 687

\bibitem[Helmi \& White(2001)]{helmi01} Helmi, A.~\& White, 
S.~D.~M.\ 2001, MNRAS, 323, 529 

\bibitem[Hernquist \& Quinn(1988)]{hernquist88} Hernquist, L.~\& Quinn, P.~J.\ 1988, ApJ, 331, 682

\bibitem[Hernquist \& Quinn(1989)]{hernquist89} Hernquist, L.~\& Quinn, P.~J.\ 1989, ApJ, 342, 1

\bibitem[Knebe et al.(2004)]{knebe04} Knebe, A., Gill, S.~P.~D., Kawata, D., \ Gibson, B.~K.\ 2004, MNRAS, submitted (astro-ph/0407418)

\bibitem[Law, Johnston, \& Majewski(2004)Law et al.]{law04} Law, D.~R., Johnston, K.~V., \& Majewski, S.~R.\ 2004, ApJ, submitted (astro-ph/0407566)

\bibitem[Malin(1978)]{malin78} Malin, D. F. 1978, Nature, 276, 591

\bibitem[Malin \& Carter(1980)]{malin80} Malin, D.~F.~\& Carter, D.\ 1980, Nature, 285, 643

\bibitem[Malin \& Carter(1983)]{malin83} Malin, D.~F.~\& Carter, D.\ 1983, ApJ, 274, 534

\bibitem[Mateo(1998)]{m98} Mateo, M. 1998, ARA\&A, 36, 435

\bibitem[Mayer et al.(2001)]{mayer01} Mayer, L., Governato, F., 
Colpi, M., Moore, B., Quinn, T., Wadsley, J., Stadel, J., \& Lake, G.\ 
2001, ApJ, 559, 754 

\bibitem[Moore et al.(1999)]{moore99} Moore, B., Ghigna, S., Governato, F., Lake, G., Quinn, T., Stadel, J., \& Tozzi, P. 1999, ApJ, 524, L19

\bibitem[Navarro, Frenk, \& White(1996)Navarro et al.]{nfw96} Navarro, J.~F., Frenk, C.~S., \& White, S.~D.~M.\ 1996, ApJ, 462, 563

\bibitem[Navarro, Frenk, \& White(1997)Navarro et al.]{nfw97} Navarro, J.~F., Frenk, C.~S., \& White, S.~D.~M.\ 1997, ApJ, 490, 493

\bibitem[Pont et al.(2004)]{pont04} Pont, F., Zinn, R., Gallart, C., Hardy, E., \& Winnick, R.\ 2004, AJ, 127, 840

\bibitem[Ratnatunga \& Bahcall(1985)]{rat85} Ratnatunga, K.~U.~\& Bahcall, J.~N.\ 1985, ApJS, 59, 63 

\bibitem[Schlegel, Finkbeiner, \& Davis(1998)Schlegel et al.]{schlegel98} Schlegel, D.~J., Finkbeiner, D.~P., \& Davis, M.\ 1998, ApJ, 500, 525 

\bibitem[Stetson, Hesser, \& Smecker-Hane(1998)Stetson et al.]{stetson98} Stetson, P. B., Hesser, J. E., \& Smecker-Hane, T. A.\ 1998, PASP, 110, 533 

\end{thebibliography}
\end{document}